\documentstyle[aps,multicol,epsfig,pstricks]{revtex}
\tighten
\begin{document}
\draft
\title{Accuracy of Sampling Quantum Phase Space
in Photon Counting Experiment} 
\author{Konrad Banaszek\cite{uw}}
\address{Optics Section, Blackett Laboratory, Imperial College,
Prince Consort Road, London SW7 2BZ, England}
\author{Krzysztof W\'{o}dkiewicz\cite{uw}}
\address{Center of Advanced Studies and Department of
Physics, University of New Mexico, Albuquerque NM 87131, USA}
\date{October 31, 1996}
\maketitle
\begin{abstract}
We study the accuracy of determining the phase space quasidistribution of
a single quantized light mode by a photon counting experiment. We
derive an exact analytical formula for the error of the experimental
outcome. This result provides an estimation for the experimental
parameters, such as the number of events, required to determine the
quasidistribution with assumed precision. Our analysis also shows that
it is in general not possible to compensate  the imperfectness of
the photodetector in a numerical processing of the experimental
data. The discussion is illustrated with Monte Carlo simulations of
the photon counting experiment
for the coherent state, the one
photon Fock state, and the Schr\"{o}dinger cat state.
\end{abstract}

\begin{multicols}{2}
\section{Introduction}
Generation and detection of simple quantum
systems exhibiting nonclassical features have been, over the past few
years, a subject of numerous experimental and theoretical studies.
Quantum optics, provides in a natural way, several interesting
examples of nonclassical  systems along with  practical tools to
perform  measurements.  
One of the simplest systems is 
a single quantized light mode, whose
Wigner function has been measured in a series of beautiful experiments
using optical homodyne tomography \cite{Tomography,BreiMullJOSA95}
following a theoretical proposal \cite{VogeRiskPRA89}.  

Recently, a novel method for the determination of the
Wigner function of a light mode has been proposed
\cite{WallVogePRA96,BanaWodkPRL96}. This
method utilizes a direct relation between the photocount
statistics and the phase space quasidistributions of light.  
Simplicity of this relation makes it possible
to probe independently each point of the single
mode phase space. In a very recent experiment, an analogous method has
been used to measure the Wigner function of the motional state of a 
trapped ion \cite{LeibMeekUNP96}. 

In the present paper we supplement the description of this method with a
rigorous analysis of the statistical uncertainty, when the phase space
quasidistribution is determined from a finite sample of measurements.
The analysis of the reconstruction errors due, for example, to a
finite set of recorded photons is essential in designing an experiment
that can provide accurate reconstruction of the phase space
quasidistribution. We will derive analytical formulae for the
uncertainty of the experimental result, which
give theoretical control over the potential sources of errors.  
Additional motivation for these studies comes from recent discussions
of the possibility of compensating the detector losses
in quantum optical measurements
\cite{DAriLeonPRA95,KissHerzPRA95}. Our analysis will provide a
detailed answer to the question whether such a compensation is possible
in the newly proposed scheme.

This paper is organized as follows. First,
in Sec.\ \ref{Sec:Quasidistributions}
we briefly review properties
of the quasidistribution functions that are relevant to the further
parts of the paper. Then, in order to make the paper
self--contained, we present the essentials of the method in 
Sec.\ \ref{Sec:Presentation}. Next we derive and discuss the
statistical error of the experimentally determined quasidistribution
in Sec.\ \ref{Sec:Error}. The general results 
are illustrated with several examples in Sec.\ \ref{Sec:Examples}. 
Finally, Sec.\ \ref{Sec:Conclusions} summarizes the paper.

\section{Quasidistributions of a single light mode}
\label{Sec:Quasidistributions}
The phase space representation of the state of a single quantized
light mode as a function, of one complex variable $\beta$, has been
extensively used in quantum 
optics since its very beginning. Due to the noncommutativity of the
boson creation and annihilation operators, the Wigner function is just
an example of   the more general  $t$-parameterized quasiprobability
distributions. Such  a one--parameter  family of quasidistribution
functions is given by the following formula \cite{CahiGlauPR69}: 
\begin{equation}
\label{Eq:QuasiDistDef}
W(\beta ; t) = \frac{1}{\pi^2} 
\int \text{d}^2 \xi \, 
e^{t |\xi|^2 /2 + \xi^\ast \beta - \xi \beta^\ast}
\left\langle e^{\xi \hat{a}^\dagger - \xi^\ast \hat{a}}
\right\rangle ,
 \end{equation}
where $\hat{a}$ and $\hat{a}^\dagger$ are  the single mode photon
annihilation and creation operators,  and the real 
parameter $t$ defines the corresponding phase space distribution. It
is well known that such a parameter is associated with the ordering of
the field  
bosonic  operators. For example, $t=1 ,0,$ and $-1$
correspond to the normal, symmetric and antinormal ordering,
respectively. The corresponding quasidistributions are: the $P$
function, the Wigner function, and the 
$Q$ function.  The properties of the $t$-parameterized
quasiprobability distributions are quite different.  For example: the
$P$ function 
is highly singular for nonclassical states of light, the Wigner
function is well behaved for all states, but may take negative values,
and finally the $Q$ function is always positive definite. These
properties 
reflect a general relation linking any two differently ordered
quasidistributions via convolution with a  Gaussian function in the
complex phase space: 
\begin{equation}
\label{QDistConvolution}
W(\beta;t') = \frac{2}{\pi(t-t')} \int \text{d}^2 \gamma \,
\exp\left( - \frac{2|\beta-\gamma|^2}{t-t'} \right)
W(\gamma;t),
\end{equation}
where $t > t'$.
Thus the lower the ordering, the smoother the quasidistribution is,
and fine details of the functions are not easily visible.

\begin{figure}
\begin{center}
\epsfig{file=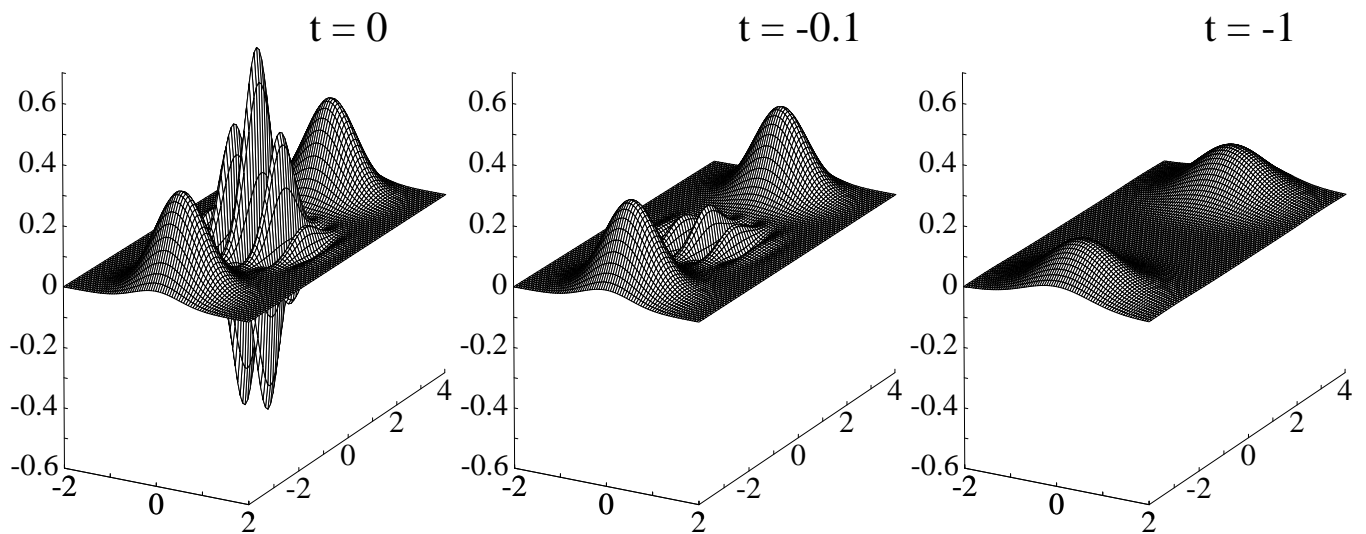,%
bbllx=80,%
bblly=296,%
bburx=525,%
bbury=472,%
width=\textwidth%
}
\end{center}
\caption{Quasidistributions representing 
the Schr\"{o}dinger cat state for
 $\alpha_0 = 3i$, depicted
for the ordering parameters $t=0, -0.1,$ and 
$-1$.\label{Fig:CatQDist}}
\end{figure}

This behavior
can be explicitly illustrated using
a superposition of two coherent states:
\begin{equation}
\label{Eq:SchroedingerCatDef}
|\psi\rangle = \frac{1}{\sqrt{2(1+e^{-2|\alpha_0|^2})}} 
(|\alpha_0\rangle + |- \alpha_0 \rangle). 
\end{equation}
States of this type illustrate quantum coherence and 
interference between classical--like components, and are often
called  
quantum optical Schr\"{o}dinger cats 
\cite{VogeRiskPRA89,SchroedingerCat}.
The quasidistribution of this superposition is given by the formula
\end{multicols}
\noindent\rule{0.5\textwidth}{0.4pt}\rule{0.4pt}{\baselineskip}
\begin{eqnarray}
W^{|\psi\rangle}(\beta; t) & = & 
\frac{1}{\pi(1-t)(1+e^{-2|\alpha_0|^2})}
\left\{
\exp\left(-\frac{2}{1-t}|\alpha_0 - \beta|^2\right) +
\exp\left(-\frac{2}{1-t}|\alpha_0 + \beta|^2\right) 
\vphantom{\left(\frac{4 \text{Im}(\alpha_0 \beta^{\ast})}{1-t} 
\right)}
\right.
\nonumber \\
& & 
\label{Eq:SchroedingerCat}
\left.
+ 2 
\exp\left(\frac{2t}{1-t}|\alpha_0|^2 - \frac{2}{1-t}|\beta|^2
\right) 
\cos \left(\frac{4 \text{Im}(\alpha_0 \beta^{\ast})}{1-t} \right)
\right\}.
\end{eqnarray}
\hspace*{\fill}\rule[0.4pt]{0.4pt}{\baselineskip}%
\rule[\baselineskip]{0.5\textwidth}{0.4pt}
\begin{multicols}{2}
\noindent
Fig.~\ref{Fig:CatQDist} shows this quasidistribution plotted for three
different values of the ordering parameter $t$. 
The Wigner function contains an oscillating component
originating from the interference between the coherent states. This
component is completely smeared out in the $Q$ function, which can
hardly be distinguished from that of a statistical mixture of two
coherent states. 

Computation of a higher ordered quasidistribution from a given one
is not a straightforward task, since the integral in
Eq.~(\ref{QDistConvolution}) fails to converge for $t'>t$. Instead, a
relation linking the Fourier transforms of the quasidistributions can
be used in analytical calculations:
\begin{equation}
\tilde{W}(\xi;t') = \exp[(t'-t)|\xi|^2/2] \tilde{W}(\xi,t). 
\end{equation}
However, its application in the processing of experimental data
would enormously amplify the statistical error \cite{LeonPaulPRL94}. 
Consequently, in the case of experimentally determined functions it is
practically
impossible to compute a higher ordered quasidistribution from the
measured one. 
Therefore the ordering of the measured quasidistribution 
depends primarily on the features of a specific experimental
scheme. Optical 
homodyne tomography is capable of
measuring the Wigner function,
provided that the detection is lossless. For 
heterodyne \cite{ShapWagnIEE84} and double homodyne \cite{WalkJMO87} 
detection schemes the $Q$ function is the limit, because in
these methods an additional noise is introduced by a 
vacuum mode mixed with the detected fields. 

In our calculations a normally ordered representation
of the quasidistribution functions will be very useful. 
Introducing normal ordering of the creation and annihilation
operators in Eq.~(\ref{Eq:QuasiDistDef})
allows to perform the integral explicitly, which yields:
\begin{equation}
\label{WbetatNormOrd}
W(\beta;t ) =
\frac{2}{\pi(1-t)} \left\langle : \exp \left( -
\frac{2}{1-t} (\hat{a}^\dagger - \beta^\ast)(\hat{a} - \beta)
\right) : \right\rangle
\end{equation}
This formula can be transformed into the form
\cite{EnglJPA89}:
\begin{equation}
\label{WbetatSymOrd}
W(\beta;t)
= 
\frac{2}{\pi(1-t)} \left\langle \left( 
\frac{t+1}{t-1} \right)^{(\hat{a}^\dagger - \beta^{\ast})
(\hat{a} - \beta)} \right\rangle
\end{equation}
showing that for $t \le 0$ the quasidistribution is an expectation
value of a bounded operator and therefore is well behaved. Of course,
only well behaved, nonsingular functions have experimental
significance. 

\section{Photocount statistics and quasidistributions}
\label{Sec:Presentation}
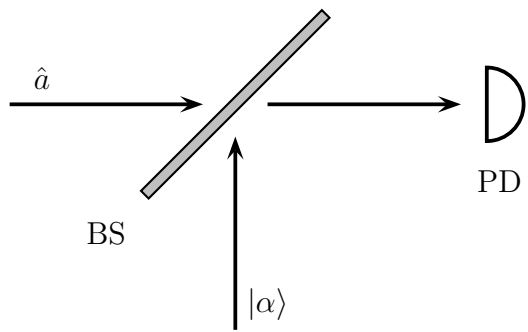
\begin{figure}
\begin{center}
\psset{unit=.3375in}
\begin{pspicture}(10,6)
\rput{45}(4.5,4){%
\psframe[fillstyle=solid,fillcolor=lightgray](-2,-.1)(2,.1)%
} 
\psset{linewidth=.5mm}
\psline{->}(1,4)(4,4)
\psline{->}(5,4)(8,4)
\psline{->}(4.5,.5)(4.5,3.5)
\pswedge(8.4,4){.6}{270}{90}
\rput(2.5,2){\large BS}
\rput(8.6,2.8){\large PD}
\rput(1.5,4.4){\large $\hat{a}$}
\rput(5.0,0.9){\large $|\alpha\rangle$}
\end{pspicture}
\end{center}
\caption{The experimental setup proposed to measure
quasidistribution functions. The signal $\hat{a}$ is superposed
by the beam splitter BS on a coherent state $|\alpha\rangle$. 
The photon statistics of this superposition
is measured by the photodetector PD. 
\label{Fig:Setup}}
\end{figure}

The experimental setup proposed to
determine the quasiprobability distribution
via photon counting is presented in Fig.~\ref{Fig:Setup}. The signal,
characterized by an annihilation operator $\hat{a}$ is superposed by a 
beam splitter on a probe field prepared
in a coherent state $|\alpha\rangle$.
The measurement is performed only on  one of the output ports of
the beam splitter, delivering the transmitted signal mixed with the
reflected probe field. The quantity detected on this port is simply
the photon statistics, measured with the help of a single
photodetector. We will consider an imperfect detector,
described by a quantum efficiency $\eta$. According to the
quantum theory of 
photodetection, the probability $p_n$ of 
registering $n$ counts by the detector 
is given by the expectation value of the 
following normally ordered operator:
\begin{equation}
p_n = \left\langle : \frac{(\eta\hat{\cal J}_{\text{out}})^n}{n!}
e^{-\eta\hat{\cal J}_{\text{out}}} :
\right\rangle,
\end{equation}
where the angular brackets denote the quantum expectation value and
 $\hat{\cal J}_{\text{out}}$ is the operator of the time--integrated 
flux of the light incident onto the surface of the detector. 
It can be expressed in terms of the signal and probe fields as
\begin{equation}
\hat{\cal J}_{\text{out}} = (\sqrt{T}\hat{a}^\dagger - 
\sqrt{1-T} \alpha^{\ast})(\sqrt{T} \hat{a} - \sqrt{1-T} \alpha), 
\end{equation}
with $T$ being the beam splitter power transmission. 
The count statistics determined in the experiment is 
used to compute the following photon count generating function (PCGF):
\begin{equation}
\label{Pialphas}
\Pi (\alpha; s)  =  \sum_{n=0}^{\infty} 
\left( \frac{s+1}{s-1} \right)^n p_n  = \left\langle : \exp \left(
- \frac{2\eta \hat{\cal J}_{\text{out}}}{1-s} \right) : \right\rangle,
\end{equation}
where $s$ is a real parameter, which will allow us to manipulate to
some extent  the ordering of the measured phase space quasidistribution. 
We will specify the  allowed range of the parameter $s$ later. 
A comparison of the latter form of the PCGF  with 
Eq.~(\ref{WbetatNormOrd}) shows that it is directly related to a
specifically ordered quasidistribution of the signal field.
After the identification of the parameters we obtain
that the PCGF can be written as:
\begin{equation}
\label{PiandQDist}
\Pi (\alpha; s) = \frac{\pi(1-s)}{2\eta T} W \left( 
\sqrt{\frac{1-T}{T}} \alpha ; - \frac{1 - s - \eta T}{\eta T}
\right).
\end{equation}
Thus the PCGF computed from the count statistics is
proportional to the value of the signal quasiprobability 
distribution at the point $\beta = \sqrt{(1-T)/T}\alpha$,
determined by the amplitude 
and the phase of the probe coherent field
\cite{MoyaKnigPRA93}. 
We may therefore scan the whole phase space of the signal 
field by changing the parameters
of the probe field. The main advantage of this
scheme is that the measurement of the quasidistribution is
performed independently at each point of the phase space. 
In contrast to optical homodyne tomography there is
no need to gather first the complete set of experimental data and then 
process it numerically.

The ordering of the measured quasidistribution depends
on two quantities: the parameter $s$ used in the construction of the
PCGF from the count statistics, and the product of 
the beam splitter transmission $T$ and the detector
efficiency $\eta$. Let us first 
consider the case of an ideal 
detector $\eta=1$ and the transmission 
tending to one. In this limit 
the ordering of the quasidistribution 
approaches $s$.  
Setting $s=0$,
which corresponds to multiplying the count statistics by $(-1)^{n}$, 
yields the Wigner function of the 
signal field. However, in the 
limit $T\rightarrow 1$ only a small fraction of the probe field is
reflected  to the detector, which results
in the vanishing factor multiplying 
$\alpha$ in Eq.~(\ref{PiandQDist}), and thus
an intense probe field has to be used to achieve the required
shift in the phase space. Consequently, 
the beam splitter should have the transmission
close to one, yet reflect enough probe field to allow scanning of the
interesting region of the phase space. This decreases the ordering of the
measured quasidistribution slightly below zero. A more
important factor that lowers the ordering is the imperfectness of the
detector. Maximum available efficiency of photodetectors 
is limited by the current technology and
on the level of single photon signals does not exceed 80\%
\cite{KwiaSteiPRA93}. Therefore in a realistic setup
the parameter deciding about the magnitude of the product $\eta T$  
is the detector efficiency rather then the beam splitter transmission.

One may attempt to compensate the lowered ordering by
setting $s= 1- \eta T$, which would allow to measure the Wigner 
function regardless of the value of $\eta T$. However in this case 
the magnitude of the factor multiplying the
count statistics in the PCGF diverges to infinity, which 
would be a source of severe problems in a real experiment. We will 
discuss this point in detail later, in the framework of a rigorous
statistical analysis of the measurement. 

Let us close this section by presenting
several examples of the photocount statistics for different quantum
states of the signal probed by a coherent source of light. 
The  most straightforward case is when a coherent state
 $|\alpha_0\rangle$ enters through the signal
port of the beam splitter. Then the statistics of the registered
counts is given by the Poisson distribution:
\begin{equation}
\label{Eq:pncoh}
p_n^{|\alpha_0\rangle} = \frac{(J(\alpha_0))^n}{n!} e^{-J(\alpha_0)},
\end{equation}
where $J(\alpha_0) = \eta T|\beta - \alpha_0|^2$ 
is the average number of registered photons. When the
measurement is performed at the point 
where the quasidistribution of the signal field is
centered, i.e., $\beta = \alpha_0$, the fields
interfere destructively and no photons are detected.
In general, for an arbitrary phase space point, the average
number of registered photons is proportional to the squared 
distance from $\alpha_0$. Averaging Eq.~(\ref{Eq:pncoh}) over
an appropriate $P$ representation yields
the photocount statistics for a thermal signal
state characterized by an average photon number $\bar{n}$:
\begin{eqnarray}
p_n^{\text{th}} & = & \frac{(\eta T \bar{n})^n}{(1+\eta T \bar{n})^{n+1}} 
L_n \left( - \frac{|\beta|^2}{\bar{n}(1+\eta T \bar{n})} 
\right)
\nonumber \\
\label{pnThermal}
& & \times
\exp \left( - \frac{\eta T |\beta|^2}{1 + \eta T \bar{n}} \right),
\end{eqnarray}
where $L_n$ denotes the $n$th Laguerre polynomial. 

A more interesting case is when the 
signal field is in a nonclassical state. Then the interference
between the signal and the probe fields cannot be described within
the classical theory of radiation. We will consider two nonclassical
states: the one photon Fock state and the Schr\"{o}dinger cat state. 
The count statistics can be obtained by calculating the quantum
expectation value of PCGF over the considered state and then expanding
it into the powers of $(s+1)/(s-1)$. 

\begin{figure}
\begin{center}
\epsfig{file=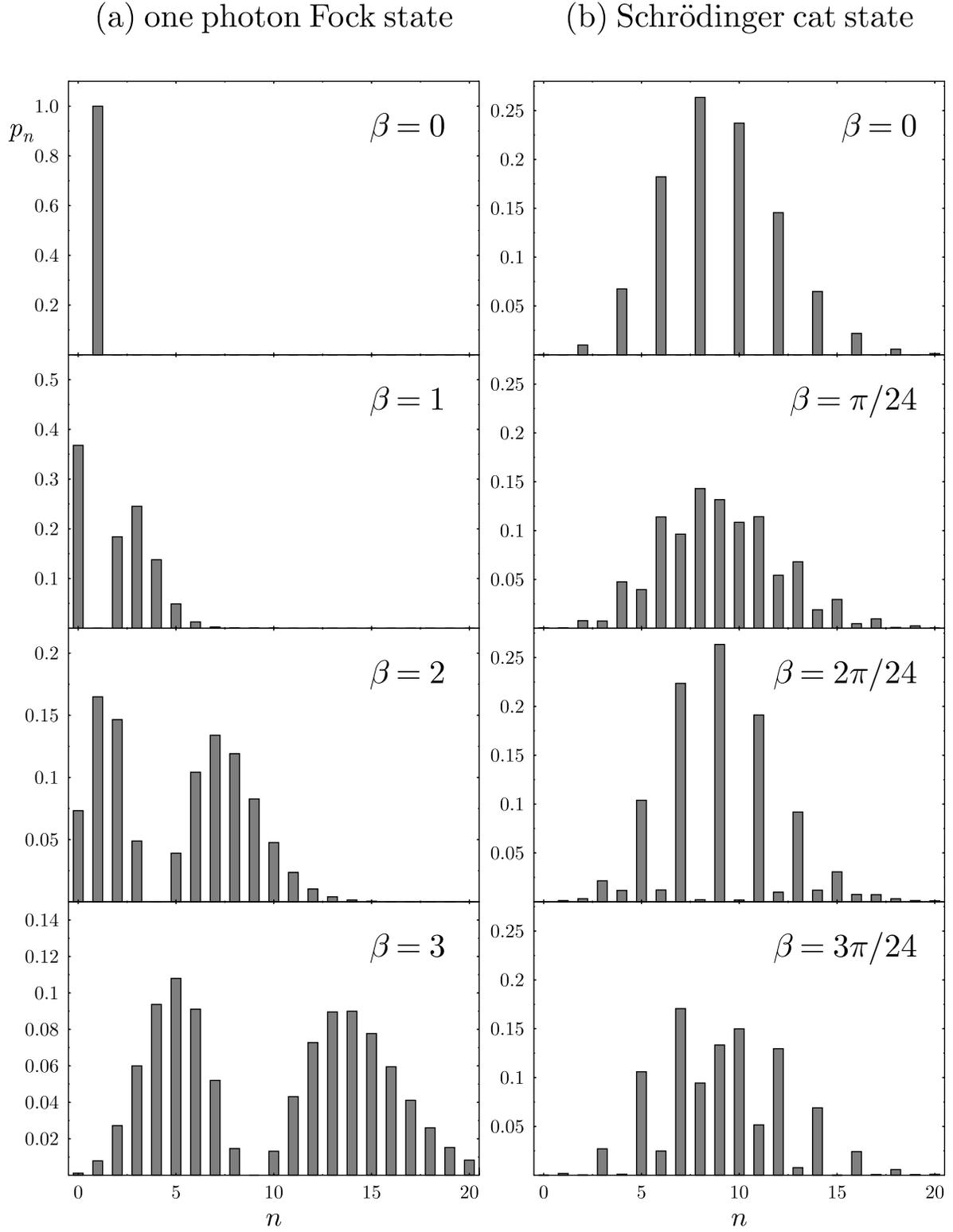,%
bbllx=45,%
bblly=55,%
bburx=545,%
bbury=710,%
width=0.9\textwidth
}
\end{center}
\caption{The photocount statistics of (a) the one photon Fock state
and (b) the Schr\"{o}dinger cat state for $\alpha_0 = 3i$,
shown for several values of the rescaled probe field amplitude 
$\beta = \protect\sqrt{(1-T)/T}\alpha$ in the limit $\eta T =1$.}
\label{Fig:Statistics}
\end{figure}

The photocount distribution for the one photon Fock 
state $|1\rangle$ can be written as an 
average of two terms with the weights $\eta T$ and $1- \eta T$:
\begin{eqnarray}
p^{|1\rangle}_n & = & \eta T (n-J(0))^2
\frac{(J(0))^{n-1}}{n!} e^{-J(0)} 
\nonumber \\
& & + (1- \eta T)
\frac{(J(0))^{n}}{n!} e^{-J(0)} \; .
\end{eqnarray}
The second term corresponds to the  detection of the vacuum signal
field. Its presence is a result of the detector imperfectness and the
leakage of the signal field through the unused output port of the beam
splitter. This term vanishes in the limit of $\eta T \rightarrow 1$, 
where the Wigner function is measured in the setup. The first term
describes the detection of the one photon Fock state. In
Fig.~\ref{Fig:Statistics}(a) we show the statistics generated
by this term for different values 
of $\beta$. If the amplitude of the probe field is
zero, we detect the undisturbed signal field and the statistics is
nonzero only for $n=1$. The distribution becomes flatter with
increasing $\beta$. Its characteristic feature is that it vanishes
around $n \approx J(0)$. 

For the Schr\"{o}dinger cat state defined in
Eq.~(\ref{Eq:SchroedingerCatDef}) the photocount
statistics is a sum of three terms:
\end{multicols}
\noindent\rule{0.5\textwidth}{0.4pt}\rule{0.4pt}{\baselineskip}
\begin{eqnarray}
p_n^{|\psi\rangle} & = &
\frac{1}{2(1+e^{-2|\alpha_0|^2})}
\left\{
\frac{(J(\alpha_0))^n}{n!} e^{-J(\alpha_0)} + 
\frac{(J(-\alpha_0))^n}{n!} e^{-J(-\alpha_0)}
\right. \nonumber \\
& & \left. + 2 {\text{Re}} \left[ 
\frac{(\eta T (\beta^{\ast} - \alpha^{\ast}_0) 
(\beta + \alpha_0))^n}{n!} 
e^{\eta T (\alpha_0^{\ast} \beta - \alpha_0 \beta^{\ast})}
\right] e ^{-(2-\eta T)|\alpha_0|^2 - \eta T |\beta|^2}
\right\} \; .
\end{eqnarray}
\hspace*{\fill}\rule[0.4pt]{0.4pt}{\baselineskip}%
\rule[\baselineskip]{0.5\textwidth}{0.4pt}
\begin{multicols}{2}
\noindent
The first two terms describe  the two coherent components of the cat
state, whereas the last one contributes to the quantum interference 
structure. In Fig.~\ref{Fig:Statistics}(b) we plot the photocount
statistics for different values
of $\beta$ probing this structure, in the 
limit $\eta T \rightarrow 1$. The four values of
$\beta$ correspond to the cosine function in
Eq.~(\ref{Eq:SchroedingerCat}) equal to $1, 0, -1,$ and $0$,
respectively, for $t=0$. 
It is seen that the form of the statistics changes very
rapidly with $\beta$. This behavior becomes clear if we rewrite
the PCGF (Eq.~(\ref{Pialphas})) for $s=0$ in the form
\begin{equation}
\Pi(\alpha; 0) = \sum_{l=0}^{\infty} p_{2l} 
- \sum_{l=0}^{\infty} p_{2l+1}
\end{equation}
showing that in order to obtain a large positive (negative) value of
the quasidistribution the photocount statistics has to be concentrated
in even (odd) values of $n$. 

\section{statistical error}
\label{Sec:Error}
The proposed measurement scheme is based on the relation between the
quasidistributions and the photocount statistics. In a real
experiment the statistics of the detector counts cannot be
known with perfect accuracy, as it is determined from a
finite sample of $N$ measurements. 
This statistical uncertainty affects the
experimental value of the quasidistribution. 
Theoretical analysis of the statistical error is important for
two reasons. First, we need an estimation for the number 
of the measurements required to determine the quasidistribution with a given
accuracy. Secondly, we have seen that one may attempt to compensate the 
imperfectness of the detector and the nonunit transmission of the beam 
splitter by manipulating the parameter $s$. However, in this case 
the magnitude of the factor multiplying the count statistics diverges
to infinity. This amplifies the contribution of the tail of the 
distribution, which is usually determined from a very small sample of data, 
and may therefore result in a huge statistical error of the final result. 
Our calculations will provide a quantitative analysis of this problem. 

Due to extreme simplicity
of Eq.~(\ref{Pialphas}), linking the count statistics with the
quasidistributions, it is possible 
to perform a rigorous analysis of the statistical error and obtain an
exact expression for the uncertainty of the final result. 
We will assume that the maximum number of
the photons that can be detected in a single measurement cannot exceed a
certain cut--off parameter $K$.  Let us denote by $k_n$
the number of measurements when $n$ photons have been detected,
 $n=0,1,\ldots,K$. The set of $k_n$'s obeys the multinomial
distribution 
\begin{eqnarray}
\label{Pk0k1}
\lefteqn{{\cal P}(k_0,k_1, \ldots , k_K )
=  \frac{N!}{k_0! k_1! \ldots 
\left(N - \sum_{n=0}^{K}k_n \right)!}}
& & \nonumber \\
& & 
\makebox[1cm]{}
\times
p_0^{k_0} p_1^{k_1} \ldots p_K^{k_K}
\left( 1 - \sum_{n=0}^{K} p_n 
\right)^{N - \sum_{n=0}^{K} k_n}.
\end{eqnarray}
The measured count statistics is converted into the experimental PCGF:
\begin{equation}
\Pi_{\text{exp}} (\alpha; s) = \frac{1}{N} \sum_{n=0}^{K} 
\left( \frac{s+1}{s-1} \right)^n k_n,
\end{equation}
which is an approximation of the series defined in 
Eq.~(\ref{Pialphas}). In order to see how well $\Pi_{\text{exp}}$
approximates the ideal quantity we will find its mean value and 
its variance averaged with respect to the distribution (\ref{Pk0k1}). 
This task is quite easy, since the only expressions we need in the
calculations are the following moments: 
\begin{eqnarray}
\overline{k_n} & = & N p_n, \nonumber \\
\overline{k_l k_n} & = & N (N-1) p_l p_n + \delta_{ln} N p_n.
\end{eqnarray}
We use the bar to denote the statistical average with respect to the
distribution ${\cal P}(k_0,\ldots k_K)$. Given this result, it is
straightforward to  
obtain:
\begin{equation}
\label{Eq:PiexpAv}
\overline{\Pi_{\text{exp}}(\alpha;s)} = \sum_{n=0}^{K} 
\left(\frac{s+1}{s-1}\right)^{n} p_n \; ,
\end{equation}
and
\begin{eqnarray}
\lefteqn{\delta\Pi_{\text{exp}}^{2}(\alpha; s) =
\overline{\left(\Pi_{\text{exp}}(\alpha;s) - 
\overline{\Pi_{\text{exp}} (\alpha;s)}\right)^2}}
& & 
\nonumber \\
& = & \frac{1}{N} \left[ \sum_{n=0}^{K} 
\left(\frac{s+1}{s-1}\right)^{2n} p_n 
- \left( \sum_{n=0}^{K} 
\left(\frac{s+1}{s-1}\right)^{n} p_n
\right)^2 
\right].
\nonumber \\
& & 
\label{Eq:PiexpDispersion}
\end{eqnarray}
The error introduced by the cut--off of the photocount statistics can
be estimated by
\begin{eqnarray}
|\overline{\Pi_{\text{exp}}(\alpha;s)} - \Pi(\alpha;s)| 
& = & 
\left| \sum_{n=K+1}^{\infty} \left( \frac{s+1}{s-1} 
\right)^n p_n \right|
\nonumber \\
& \le &
\sum_{n=K+1}^{\infty} 
\left|\frac{s+1}{s-1}\right|^{n} p_n.
\end{eqnarray}
The variance $\delta\Pi^{2}_{\text{exp}}$
is a difference of two terms. 
The second one is simply the squared average of $\Pi_{\text{exp}}$.
The first term is  a sum over the count statistics
multiplied by the powers of a {\it positive} factor 
$((s+1)(s-1))^2$. If $s>0$, this factor is greater than one and 
the sum may be arbitrarily large. In the case when the contribution
from the cut tail of the statistics is negligible, i.e., if 
$K \rightarrow \infty$, it can be estimated by the average number of
registered photons:
\begin{equation}
\sum_{n=0}^{\infty} \left(
\frac{s+1}{s-1} \right)^{2n} p_n \ge 1 + \frac{4 s}{(s-1)^2}
\langle \eta \hat{\cal J}_{\text{out}} \rangle.
\end{equation}
Thus, the variance grows unlimited as we probe phase space points
far from the area 
where the quasidistribution is localized. Several examples in the next
section will demonstrate that the variance usually explodes much more 
rapidly, exponentially rather than linearly. This makes the
compensation of the detector inefficiency a very subtle matter. It can
be successful only for very restricted regions of the phase space,
where the count statistics is concentrated for a small number of counts 
and vanishes sufficiently quickly for larger $n$'s. 

Therefore, in order to ensure that the statistical error remains bounded 
over the whole phase space, we have to impose the condition $s\le 0$.
Since we are interested in achieving the highest possible ordering 
of the measured
quasidistribution, we should consequently set $s=0$. For this particular
value the estimations for the uncertainty of $\Pi_{\text{exp}}$
take a much simpler form. The error 
caused by the cut--off of the count distribution can be estimated 
by the ``lacking'' part of the probability:
\begin{equation}
\label{Eq:CutOffError}
|\Pi_{\text{exp}}(\alpha;0) - \Pi (\alpha;0)|
\le 1 - \sum_{n=0}^{K} p_n,
\end{equation}
which shows that the cut--off is unimportant as long as the probability
of registering more than $K$ photons is negligible. The variance of 
$\Pi_{\text{exp}}$ is given by
\begin{eqnarray}
\delta \Pi_{\text{exp}}^{2}(\alpha;0)
& = & \frac{1}{N}\left[
\sum_{n=0}^{K} p_k - 
\left(
\overline{\Pi_{\text{exp}}(\alpha;0)}
\right)^2
\right]
\nonumber \\
& \le &
\frac{1}{N}\left[1 - \left(
\overline{\Pi_{\text{exp}}(\alpha;0)}
\right)^2
\right]
\le 
\frac{1}{N}.
\end{eqnarray}
Thus, the statistical uncertainty of the measured quasidistribution
can be simply estimated as $1/\sqrt{N}$
multiplied by the proportionality constant given 
in Eq.~(\ref{PiandQDist}). It is also seen that the uncertainty is
smaller for the phase space points where the 
magnitude of the quasidistribution is large. 

Analysis of the statistical error for the recently demonstrated
measurement of the Wigner function of a trapped ion is different,
as the analog of $p_n$ is not determined by a counting--like method
\cite{MeekMonrPRL96}. A
quantity that is detected in the experimental setup is the
time--dependence of the fluorescence light produced by driving a selected
transition. This signal is a linear combination of the populations
of the trap levels with known time--dependent
coefficients, and the $p_n$'s are 
extracted from these data by solving the resulting
overdetermined set of linear equations. Consequently, their
statistical error is not described by a simple
analytical formula. Additional source of error is the uncertainty of
the phase space displacement, which 
is performed by applying an rf field. An
element of our analysis that can be directly transferred to the case
of a trapped ion is the effect of the cut--off of the count
statistics.

\section{Examples}
\label{Sec:Examples}

We will now consider several examples of the reconstruction of the
quasidistributions from the data collected in a photon counting
experiment. Our discussion will be based on Monte Carlo simulations
compared with the analytical results obtained in the previous
section. 

First, let us note that the huge statistical error is not the only
problem in compensating the detector inefficiency. If $s>0$, the sum
(\ref{Eq:PiexpAv})  does not even have to converge in the limit of
$K\rightarrow \infty$. An example of this pathological behavior is
provided by a thermal state, which has been calculated in
Eq.~(\ref{pnThermal}). For the zero probe field we obtain 
\begin{equation}
\overline{\Pi_{\text{exp}}^{\text{th}}(0;s)} 
= \frac{1}{1+\eta T \bar{n}} 
\sum_{n=0}^{K} 
\left(
\frac{s+1}{s-1}
\right)^{n}
\left(
\frac{\eta T \bar{n}}{1 + \eta T \bar{n}}
\right)^{n},
\end{equation}
which shows that if $s>0$, then for 
a sufficiently intense thermal state
the magnitude of the summand is larger than one and consequently the
sum diverges, when $K \rightarrow \infty$. This behavior is due to the
very slowly vanishing count distribution and  it does not appear for
the other examples of the count statistics derived in
Sec.~\ref{Sec:Presentation}. 

\begin{figure}
\begin{center}
\epsfig{file=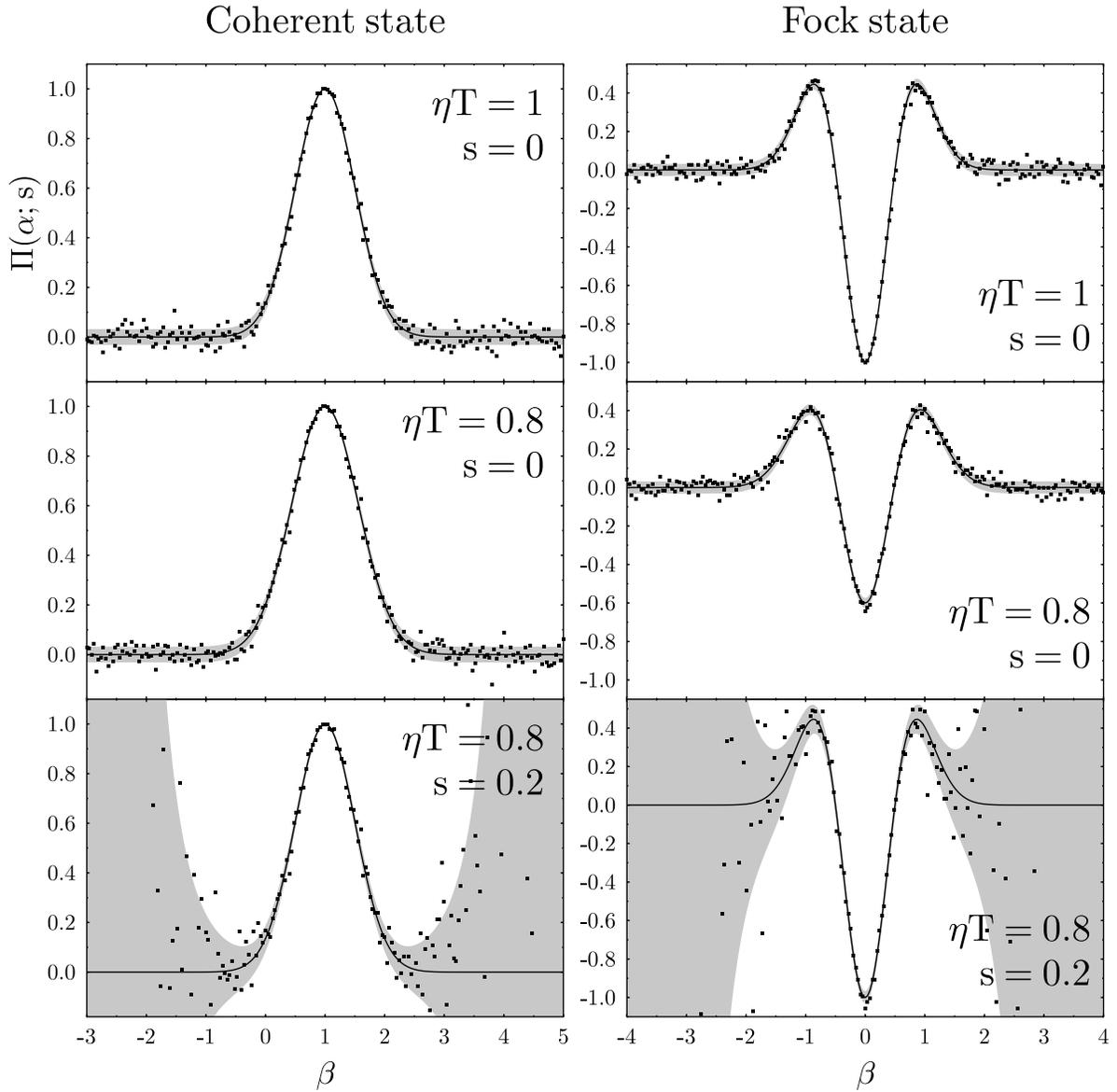,%
bbllx=45,%
bblly=55,%
bburx=550,%
bbury=550,%
width=0.9\textwidth%
}
\end{center}
\caption{Reconstruction of the quasiprobability distributions of the
coherent state $|\alpha_0=1\rangle$ (left) and the one photon Fock
state (right) from $N=1000$ events. The solid lines are the analytical
quasidistributions and the grey areas mark the statistical dispersion.}
\label{Fig:CohAndFockReconstruction}
\end{figure}

In Fig.~\ref{Fig:CohAndFockReconstruction} we plot the
reconstructed quasidistributions for the coherent state 
\mbox{$|\alpha_0 = 1\rangle$} 
and the one photon Fock state. Due to the symmetry of these
states, it is sufficient to discuss the behavior of the reconstructed
quasidistribution on the real axis of the phase space. The cut--off
parameter is set high enough to make the contribution
from the cut tail of the statistics negligibly small.
The quasidistributions are
determined at each phase space point from the Monte Carlo simulations
of $N=1000$ events. The grey areas denote the statistical uncertainty
calculated 
according to Eq.~(\ref{Eq:PiexpDispersion}). 
The two top graphs show the reconstruction of the Wigner
function in the ideal case $\eta T = 1$. 
It is seen that the statistical error is
smaller, where the magnitude of the Wigner function is large. In the
outer regions it approaches its maximum value $1/\sqrt{N}$. The effect
of the nonunit $\eta T$ is shown in the center graphs. The measured
quasidistributions become wider and the negative dip in the case of
the Fock state is shallower. In the bottom graphs we depict the
result of compensating the nonunit value of $\eta T$ by setting
 $s=1-\eta T$. The compensation works quite well in the central region,
where the average number of detected photons is small, but
outside of this region the statistical error explodes exponentially. 
Of course, the statistical error can be suppressed by increasing the
number of measurements. However, this is not a practical method, since
the statistical error decreases with the size of the sample only as 
 $1/\sqrt{N}$. 

\begin{figure}
\begin{center}
\epsfig{file=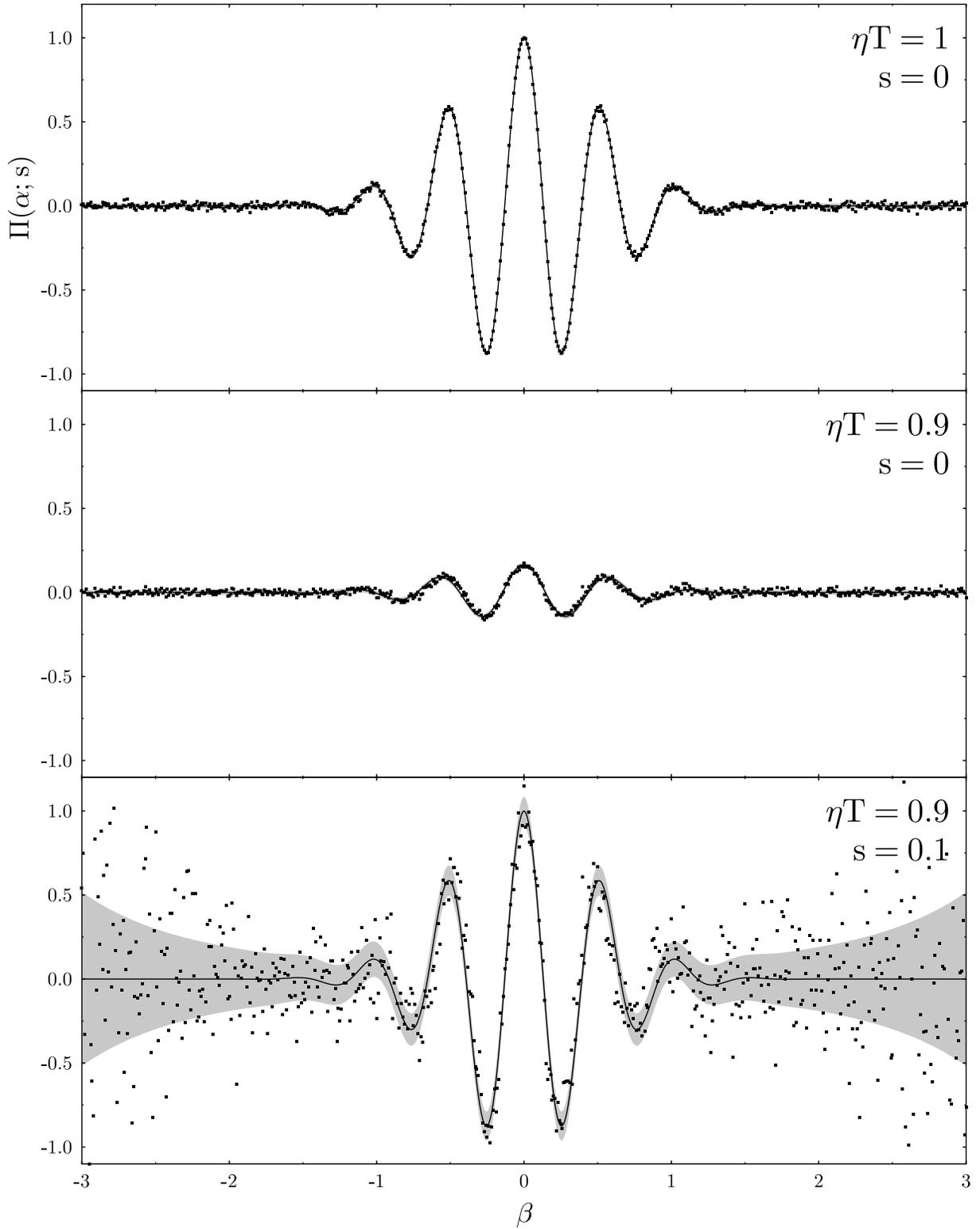,%
bbllx=45,%
bblly=50,%
bburx=540,%
bbury=680,%
width=0.9\textwidth%
}
\end{center}
\caption{Reconstruction of the interference structure of the
Schr\"{o}dinger cat state for $\alpha_0 = 3i$ from $N=5000$ events at
each point.}
\label{Fig:CatReconstruction}
\end{figure}

The reconstruction of the interference structure of the
Schr\"{o}dinger cat state is plotted in
Fig.~\ref{Fig:CatReconstruction}. This structure is very fragile, and
its precise measurement requires a large sample of events. In the
case of the presented plot,  $N = 5000$ simulations were performed at
each phase space point. Comparison of the top and the center graphs
shows how even  
a small imperfectness in the detection destroys the interference pattern. 
The data collected in an nonideal setup can be processed to recover
the Wigner function, but at the cost of a significantly larger
statistical error, as it is shown in the bottom graph. Outside the
interference structure, we again observe the exponential explosion of
the dispersion due to the increasing intensity of the detected light. 

\begin{figure}
\begin{center}
\epsfig{file=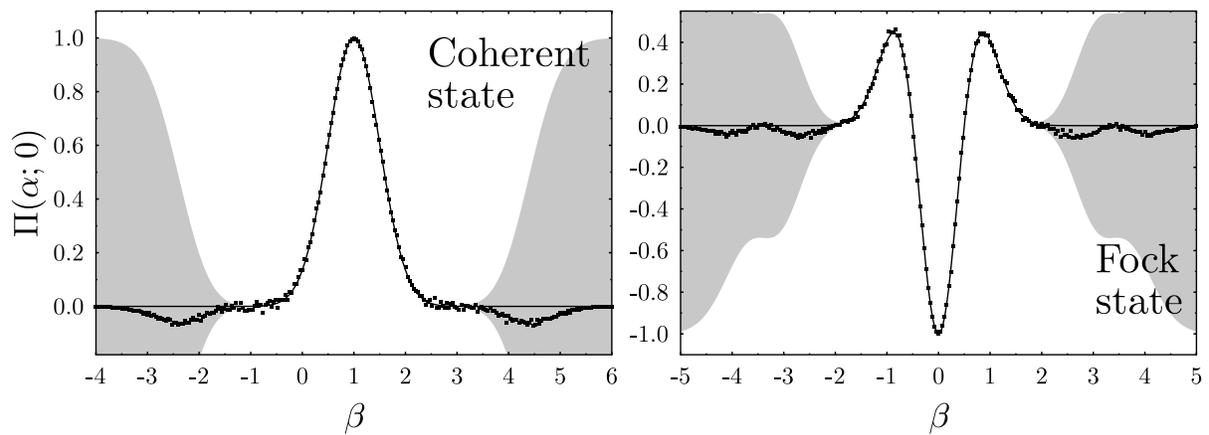,%
bbllx=45,%
bblly=50,%
bburx=545,%
bbury=230,%
width=0.9\textwidth
}
\end{center}
\caption{Reconstruction of the Wigner function of the coherent state
and the one photon Fock state from the count statistics cut at
$K=11$, for $\eta T =1$ and $s=0$. The number of events is $N=10^4$.}
\label{Fig:CutOff}
\end{figure}

Finally, let us look at the effect of cutting the statistics at a finite
value. Fig.~\ref{Fig:CutOff} shows the Wigner functions for the
one photon coherent and Fock states reconstructed from the count
distributions cut at $K=11$. We performed a large number of $N=10^4$
simulations in order to get the pure effect of the cut--off that is
not spoiled by the statistical uncertainty. The grey areas show the
cut--off error, estimated using
Eq.~(\ref{Eq:CutOffError}). The
reconstruction works well as long as the probability of detecting
more than $K$ photons is negligible. When the average number of
incident photons starts to be comparable with the cut--off parameter,
``ghost'' peaks appear. When we go even further, the Wigner function
again tends to zero, 
but this is a purely artificial effect due to the virtually vanishing
count distribution below $K$. 

\section{Conclusions}
\label{Sec:Conclusions}
The newly proposed scheme for measuring the quasiprobability
distributions is based on a much more direct link with the data
collected in an experiment than optical homodyne tomography. 
Furthermore, each point of the phase space can be probed
independently. The simplicity of the underlying idea allowed us to discuss
rigorously the experimental issues, such us the statistical uncertainty
and the effect of the cut--off of the count statistics. 

We have seen that it is, in general, not possible to compensate the
detector inefficiency. This conclusion may seem to be surprising, since it
was shown  that the true photon distribution can be
reconstructed from photocount statistics of an imperfect detector,
provided that its efficiency is larger that 50\%
\cite{KissHerzPRA95}. Moreover, a simple
calculation shows that an application of this recipe for recovering the
photon distribution is equivalent to setting the
parameter $s$ above zero. The solution of this seeming contradiction is
easy: the proposed recipe can be safely used as long as we are
interested in the photon statistics itself. However, the statistical
errors of the reconstructed probabilities are not independent, 
and an attempt to utilize these probabilities to 
calculate another physical quantity may lead
to accumulation of the statistical 
fluctuations and generate a useless
result. The simple alternating
series $\sum_{n} (-1)^{n} p_{n}$ is an
example of such a quantity. 

\section*{Acknowledgments}
This work has been partially supported 
by the Polish KBN grant 2 PO3B 006 11.
K.B. thanks the European Physical
Society for the EPS/SOROS
Mobility Grant.

\end{multicols}

\pagebreak

\vspace*{\fill}

\vspace*{\fill}

\pagebreak

\vspace*{\fill}

\vspace*{\fill}

\pagebreak

\vspace*{\fill}

\vspace*{\fill}

\pagebreak

\vspace*{\fill}

\vspace*{\fill}

\end{document}